\documentclass[conference]{IEEEtran}
\IEEEoverridecommandlockouts
\usepackage{cite}
\usepackage{amsmath,amssymb,amsfonts}
\usepackage{algorithmic}
\usepackage{graphicx}
\usepackage{textcomp}
\usepackage{multirow}
\usepackage{enumitem}
\usepackage{caption}
\usepackage{subcaption}
\usepackage{comment}
\usepackage{booktabs}
\usepackage[table,xcdraw]{xcolor}
\providecommand{\keywords}[1]{\textbf{\textit{Index terms---}} #1}
\def\BibTeX{{\rm B\kern-.05em{\sc i\kern-.025em b}\kern-.08em
    T\kern-.1667em\lower.7ex\hbox{E}\kern-.125emX}}
\begin{document}

\title{Demonstration of Safe Electromagnetic Radiation Emitted by 5G Active Antenna Systems}
\author
{
\IEEEauthorblockN
{
Sumit Kumar, Chandan Kumar Sheemar,
Abdelrahman Astro, 
Jorge Querol,
Symeon Chatzinotas
} \vspace{-3mm}
\\
\IEEEauthorblockA{Interdisciplinary Centre for Security, Reliability and Trust (SnT), University of Luxembourg, Luxembourg} \vspace{-3mm}
\\
Email:\{sumit.kumar, chandankumar.sheemar jorge.querol,symeon.chatzinotas\}@uni.lu

}
\maketitle
\begin{abstract}
The careful planning and safe deployment of 5G technologies will bring enormous benefits to society and economy. Higher frequency, beamforming and small-cells are key technologies that will provide unmatched throughput and seamless connectivity to the 5G users. Superficial knowledge of these technologies has raised concerns among the general public about the harmful effects of radiation. Several standardization bodies are active to put limits on the emissions which are based on a defined set of radiation measurement methodologies. However, due to the peculiarity of 5G such as dynamicity of the beams, network densification, Time Division Duplexing mode of operation, etc, using existing EMF measurement methods may provide inaccurate results. 
In this context, we discuss our experimental studies aimed towards the measurement of radiation caused by beam-based transmissions from 5G base-station equipped with an Active Antenna System(AAS). We elaborate on the shortcomings of current measurement methodologies and address several open questions. Next, we demonstrate that using user-specific downlink beamforming, not only better performance is achieved compared to non-beamformed downlink, but also the radiation in the vicinity of the intended user is significantly decreased. Further, we show that under weak reception conditions, an uplink transmission can cause significantly high radiation in the vicinity of the user-equipment. We believe that our work will help in clearing several misleading concepts about the 5G EMF radiation effects. We conclude the work by providing guidelines to improve the methodology of EMF measurement by considering the spatio-temporal dynamicity of the 5G transmission. 
\end{abstract}
\maketitle
\section{Introduction}\label{sec:introduction}
Fifth-generation (5G) New-Radio (NR) mobile networks have become the foundation for industrial transformation in terms of digitalization and advanced communication systems \cite{Wu}. Thanks to the technologies such as Massive MIMO, Millimeter Wave (mmWave), Beamforming, D2D, and Ultra-Dense Networks, 5G-NR can support higher data rates (upto $10~$Gbps) and significantly lower latency(maximum $10~$ms) than the previous generation 4G Long Term Evolution (LTE)\cite{Falahy, fager2019linearity}. The dense deployment of the 5G base stations is crucial for achieving the "everywhere/anytime" connectivity promised by 5G. This deployment will be on top of the existing 2G/3G/4G deployments, which are already operational. Thus, foreseeing such an increased number of active radiating nodes, a large part of the population is concerned about the potential health impacts due to increased electromagnetic field (EMF) exposure \cite{Frank,Chiaraviglio}. 

One of the particular features of 5G-NR is the application of Active Antenna Systems (AASs) which creates and manages user-specific beams. This feature helps to overcome the path loss for synchronization signals and data channels. AAS is the key enabler for the technologies such as massive MIMO, mmWave, beamforming, beam-steering \cite{fager2019linearity}. In AAS, active components from the Remote Radio Unit (RRU) are integrated with the passive antenna\cite{lagunas20205g}. This allows AAS to both amplify and convert radio frequency waves in addition to signal radiation and reception. Moreover, an integrated RRU facilitates focusing the power in a certain direction and creates \textit{Field Strength Hotspots} in the main lobes of the beams, resulting in increased SNR in the intended direction. However, such \textit{Field Strength Hotspots} has created concern among the general public over increased radiation.  

Some of the concerns from the general public include the risk of cancer, increased sensitivity to radiation, and oxidative stress in the cells\cite{pecoraro2022biological,di2018towards,kostoff2020adverse}. Such doubts and concerns can delay the deployment of 5G in some countries, leading to unfavorable economic impacts\cite{jones2020switzerland}. Standardization bodies such as International Commission on Non-Ionizing Radiation Protection (ICNIRP), International Electrotechnical Commission (IEC), and International Telecommunication Union (ITU), based on various measurement campaigns, studies, and scientific efforts, define the limits of EMF radiation. They impose that the highest possible power and gain values are used when calculating the RF-EMF exposure to ensure conservative results. Additionally, countries can follow the guidelines or even set stricter (more conservative) limits on radiation. However, such approaches can have a direct impact on the system capacity, and coverage \cite{Zhao}. A conservative limit imposed by the countries can restrict the buildout of 5G network capacity. Besides, the current EMF measurement methodologies could be challenging due to the spatial and temporal dynamics of the beams as the methodologies were developed for channels and signals consisting of \textit{always-on} pilots and \textit{non-beamformed} signals. 

In a typical operation of 5G, the signals and channels are beam-formed and the beams could be dynamic in nature, i.e., the beamwidth can vary and the beams may move to focus on the mobile UE, for example.  
Therefore, it is crucial to analyze the limitations of the current EMF assessment methodologies, study the peculiarities of 5G-NR, which pose a challenge to the current methods and identify the gaps and propose appropriate adaptations. 





\subsection{Previous Works}
The research for EMF exposure assessment in 5G cellular networks is ongoing, and several new methods are  proposed \cite{elzanaty20215g,chiaraviglio2018planning,thors2016exposure,colombi2015implications,lee2021emf,colombi2020analysis,migliore2020power}. In \cite{thors2016exposure}, the authors conducted a study to investigate the maximum transmitted power and maximum Effective Isotropic Radiated Power (EIRP) to comply with all major RF-EMF exposure standards. The power density was evaluated, and the spatial peak power density was assessed using a minimum spatial sampling density of four samples per wavelength. In \cite{colombi2015implications}, the impact of the maximum radiated power from a device used near the human body is investigated. Assessments were made using numerical simulations with the commercial electromagnetic solver FEKO. In \cite{lee2021emf}, the transmitted power of the uplink slots and the synchronization signal reference signal received power (SS-RSRP) from user equipment (UE) were measured in Seoul, Korea, to evaluate the EMF exposure. The power samples were averaged over a $1$s duration, and the measurement results showed that the time-average level, when exposed to a beam sweep of the base stations, was less than $5~\mu W/m^2$. In \cite{colombi2020analysis}, EMF exposure for several 5G base stations was measured in Australia. The base stations used the massive MIMO antennas with beamforming to optimize the signal strength at the user’s
device, and the maximum time-averaged power per beam direction was found to be well below the theoretical maximum. In \cite{migliore2020power}, a method for power reduction estimation in realistic scenarios with 5G AAS to enable a preliminary fast estimation of the EMF exposure in 5G cells is proposed.

\subsection{Main Contributions}




In this context of EMF measurement and concerns from the general public, we discuss the studies and experimentation conducted with the aim to showcase the benefits of 5G from a scientific point of view and to expose some of the misconceptions related to AAS technology. 
In particular, our work focuses on demonstrating that by using beamforming in an AAS, the SNR at the intended user is improved and at the same time EMF level in the vicinity of the intended user reduces significantly. Our main contributions to this work are summarized as follows: 
\begin{itemize}
\item We have studied the peculiarities of 5G AAS, on both the Radio Access Network and Network layer, which makes the conventional EMF measurement methods less efficient. We have identified the issues with conventional EMF measurement methodology while measuring EMF from a 5G AAS. 
\item Further, we conducted field experiments to validate the fact that a 5G AAS is capable of not only improving the SNR at the intended user but can also reducing the EMF radiation in the vicinity of the intended user. Moreover, we also studied the uplink traffic, which has gained lesser attention compared to downlink traffic, however, from an EMF radiation point of view, it is more critical to analyze. 
\item Finally, we discuss the limitation of the exiting EMF assessment methods and why they cannot be used for the higher frequency bands. To overcome the drawbacks, we discuss a novel research direction which is particularly suited for mmWave and beyond.
\end{itemize}

\emph{Paper Organization:} The rest of this paper is organized as follows. Section-\ref{sec:5g-emf-concerns} discusses the features of 5G-NR which are creating concerns over EMF radiation. Section-\ref{sec:5g-emf-meas-chan} discusses the measurement challenges of conventional EMF measurement methodologies and identifies the gaps. Section-\ref{sec:field-exp} details the field experiments conducted to validate some of the methods discussed in Section-\ref{sec:5g-emf-meas-chan} for downlink and uplink. Next, in Section-\ref{sec:open-challenges}, we discuss open challenges in 5G EMF measurement especially due to the spatio-temporal dynamicity of 5G transmissions and provide guidelines for the development of EMF measurement methodology. Finally, Section-\ref{sec:conclusions} summarizes the paper with concluding remarks on future works. 

 
\section{5G-NR EMF Concerns}\label{sec:5g-emf-concerns}

Although 5G-NR still uses OFDM for uplink and downlink, it has gone through significant changes on Radio Access Network (RAN); especially how the waveforms are transmitted/received from the gNBs. Apart from different frequency bands (FR1 and FR2), wider bandwidths ($100$ MHz for FR1 and $400$ MHz for FR2) and Time Division Duplexing (TDD) are some of the peculiarities in 5G which are in contrast to the RAN of 4G LTE. Discussion on such differences between 4G and 5G is critical from an EMF radiation and assessment point of view. In this section, we focus on the specific features of 5G-NR, which are creating concern among the general public. 

\subsection{Massive MIMO}\label{sub-massive-mimo}
Massive MIMO is one of the most important 5G technologies \cite{chataut2020massive}. It contains a large number of antenna elements on the base-station side. Through digital and/or analog control of the RF chains, one or many user-centric sharp beams can be created and the beams can be further adapted dynamically based on the user movement and requirements. A controller is required for such operation. The controller can be either collocated with the antenna units or connected via Common Public Radio Interface (CPRI). The former constitutes an Active Antenna System (AAS) which has become the dominant configuration in the ongoing deployment of 5G base stations. 

Superficially, one can correspond a large number of antenna elements to an increased amount of radiated power which in turn can strengthen the myth about higher radiation from a 5G base station. In fact, the maximum transmitted power by a 5G BS can reach up to $200$Watts which is almost double the corresponding value for a 4G BS\cite{elzanaty20215g}. This increase in power triggers the population’s concern about potential health risks. However, scientifically this is not the fact. A Massive MIMO system aims to project focused power towards user locations and in this process, compared to a sectoral antenna, less power per antenna element is required in order to achieve the \textit{same SNR} a sectoral-antenna would have provided \cite{chataut2020massive}. This is possible because the antenna array can concentrate the input energy in the desired direction and minimizes the radiation in the undesired direction, thus less input energy is required per antenna element.

\subsection{Beamformed Channels and Signals}
5G inherently uses beamforming using AAS as discussed in section-\ref{sub-massive-mimo}. Not only the user-centric data(payload and control) is beamformed, but the synchronization signals, which are used to set up the time-frequency alignment, are also transmitted as beams. The former is called \textit{User-Specific Beamforming} and is done for PDSCH and PDCCH while the latter is called \textit{Synchronization Signal Blocks (SSBs) Beam-Sweeping} which is done for PSS, SSS, and PBCH. User-specific beams could be narrower than the beams used for SSBs. Besides, the beams can be parsed in both azimuth and elevation with the application of an AAS. 

Such beam-based operations are also one of the causes of public concerns regarding radiation from the 5G base station. A crude perception of the beam, among the general public is seen as a focussed beam of energy towards them, and hence beams are visualized as a cause of concern. However, the reality is different. For a desired SNR at the intended user, the transmit power from a sectoral antenna is higher compared to an antenna array with high gain and directivity. Besides, the radiation from the sectoral antenna not only reaches the intended user but also gets spread in the vicinity unnecessarily. In contrast, the antenna array elements inside the AAS, require less input power per antenna element for a given target SNR at the intended user and at the same time minimize the spill of radiation in the vicinity of the intended user.   
\subsection{mmWave and network densification}
Millimeter-wave (mmWave) frequency bands contain large signal bandwidth and are very promising for 5G to reach a data rate as high as $10$Gb/s. Mm-Wave provides a much larger bandwidth, higher throughput, faster data rate, and capacity compared with, 3G, and 4G. Hence, 5G networks are also planned to be deployed in FR2 (for higher bandwidth use-cases) apart from FR1 ( more favorable for coverage). However, the benefits of mmWave come at the cost of reduced coverage. mmWave cannot travel as far as the frequencies used in 2G/3G/4G can travel. To cope-up with this, either beamforming is done on the mmWave carrier (discussed in section-\ref{sub-massive-mimo}) and/or the base stations have to be placed densely. The latter is called network densification and this has raised concern among the general public where network densification is being perceived as radiation densification. 
\subsection{small-cells}
Among one of the promises of 5G is a reliable and ultra-low latency connection. For this purpose it is planned to densify the deployment of 5G base-stations which is in contrast to 3G and 4G where the distance between the towers was in order of KMs. 
As the distance between the users and the nearest cell shrinks this has increased the general public's perception of potential health impacts. However, in reality, and unlikely the common perception, network densification can significantly reduce the average EMF exposure for a given required rate \cite{chiaraviglio2020will, Chiaraviglio-web}. In fact, in a small cell,  the cell coverage area is smaller for networks with higher BS density, requiring lower transmitted power. In fact, 3G and 4G with distant base-stations, radiate more power in order to cover a large chunk of area. This also leads to higher exposure of radiation at the cell egde. In contrast, in a small-cell there is a homogeneous distribution of power over the coverage area leading to uniform distribution of EMF. Densifying the network with small-cells in fact reduces the average EMF exposure over the coverage area compared to the sparsely deployed legacy networks as reported in \cite{chiaraviglio2020will,chiaraviglio20205g}.

\subsection{High Uplink Traffic}\label{subsec-high-uplink-traffic}
Traffic patterns are changing. Applications such as online gaming, cloud storage, personal broadcasting are dominated by uplink traffic. TDD based operations are more flexible compared to FDD in terms of spectrum usage. That is why TDD has become the dominant deployment mode in 5G. TDD uses the same frequency, though different time slots for uplink and downlink traffic. These time slots are configurable due to which the bandwidth demand of uplink or downlink can be catered dynamically. 


A rise in such uplink traffic is a genuine concern from the EMF point of view, especially when the RSSI at the UE is poor \cite{zaidi20185g}.   

\section{5G-NR EMF Measurement Challenges}\label{sec:5g-emf-meas-chan}
5G is definitely more challenging when it comes to measuring the EMF radiation. For classic base station antennas with static antenna radiation patterns, the RF EMF evaluations are relatively straightforward. This is completely in contrast to how 5G is designed and operated. Besides, the measurement methodologies and the maximum level set for EMF radiation vary from country to country. Regulations typically require that the highest possible power and gain values are used when calculating the RF-EMF exposure to ensure conservative results. In this section, we will discuss the peculiar features of 5G which can make the measurement of EMF difficult in comparison to 2G/3G/4G. Such features not only include the significantly different frame structure, Massive MIMO, and beam-based operations in 5G-NR but also the way in which 5G is operated on a network level. A majority of the challenges come because some of the transmitter parameters are not known apriori. 

\subsection{No Always-On Signal}
The basic principle of assessing the EMF is to measure the power received from a constant radio frequency source, which is typically a pilot signal coming from the base station and to apply a proper extrapolation factor. This kind of approach is standardized for 2G, 3G, and 4G where always-on reference signals were used as reference signals for precise channel estimation and cell-specific synchronization. However, this is not straightforward for 5G, because there is not always-on signal in 5G. Besides, 5G NR is a completely new approach regarding cell-specific signals \cite{zaidi20185g}. Unlike LTE, the reference signals in 5G (DMRS, PTRS, SRS, and CSI-RS) are transmitted only when necessary and not spread over the entire spectrum. In 5G NR only a minimum amount of cell-specific signals are broadcasted that can be measured by a calibrated measurement receiver. Rest of the signals are specific to UE and their appearance in the frequency and time domain depends on the data traffic. The only always-on signal is the synchronization signal block (SSB) which is sparse in both time and frequency. Thus it becomes difficult to extrapolate the power measured from SSB to the traffic region of the resource gird. Further, the power gain offset between SSS and PDSCH is required for extrapolation which can only be obtained with an active connection between the gNB and the scanner \cite{franci2020experimental, adda2020theoretical}. Besides, since the SSB beams are narrower, several such beams (with a unique directional index) are transmitted periodically to cover the entire sector. It is difficult to get enough dwell time to measure, especially in the case of mmWave where the number of such beams could be very high. 
 \begin{figure}[ht]
	\centering
	\includegraphics[width=0.9\linewidth]{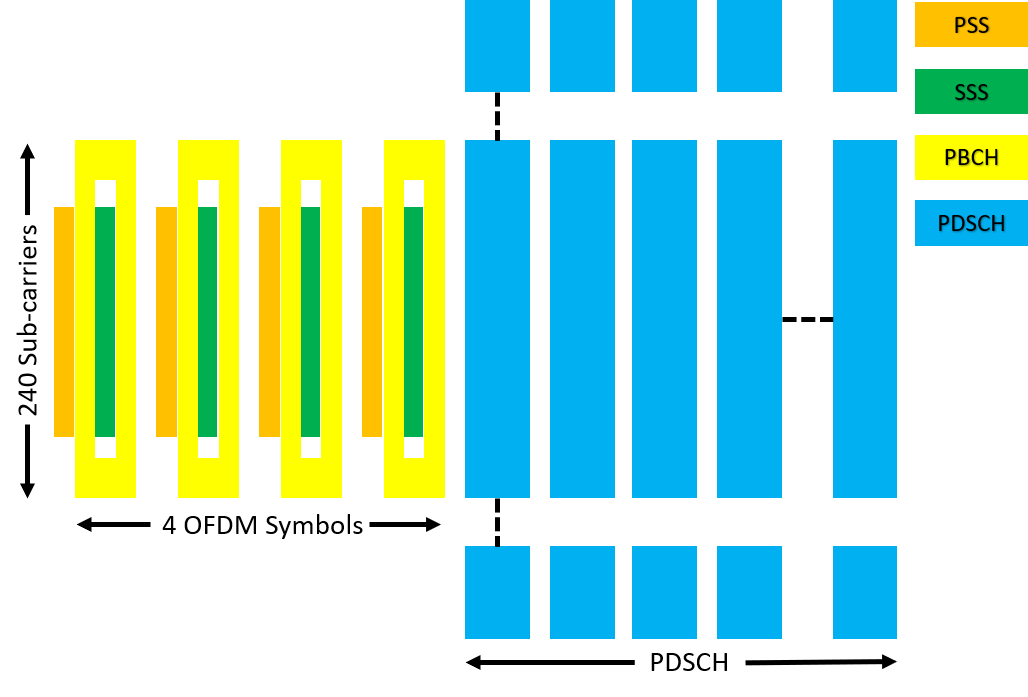}
	\caption{5G-NR Synchronization Signal Block (PSS< SSS and PBCH) and PDSCH. SSB is restricted only to 240 subcarriersin frequency domain and only 4 OFDM symbols in the time domain.}
	\label{fig:5g-resource-grid}
\end{figure}


\subsection{Traffic load distribution}
\begin{figure}[h]
	\centering
	\includegraphics[width=\linewidth]{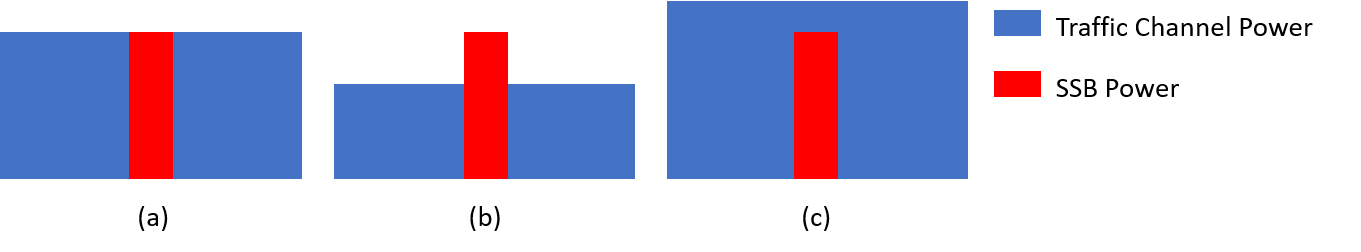}
	\caption{(a) A fully loaded cell where traffic channel power is the same as SSB power; (b) A partially loaded cell where traffic channel power is less than SSB power (c) A fully loaded cell where traffic channel power is more than SSB power}
	\label{fig:traffic-channel-data-channel-power}
\end{figure}

Since SSB is used for EMF measurement, it is important to be aware of the traffic load distribution in the resource grid. It is not straightforward to access the traffic load distribution and hence the power from the traffic channel. The reason being 5G is packet-switched, thus traffic channel power is only measurable when there is active traffic. Without the knowledge of such distribution, the linear spectral extrapolation from SSB to total channel bandwidth power does not work correctly. There could be the following three cases (a pictorial representation is shown in Figure-\ref{fig:traffic-channel-data-channel-power}): 
\begin{enumerate}[label=(\alph*)]
	\item \textit{\textbf{Cell is fully loaded, i.e., traffic channel power = SSB power }}\textbf{: }In this case, a linear spectral extrapolation from SSB will work correctly, EMF value derived from SS-RSRP will correctly represent the real power.
	\item \textit{\textbf{Cell is partially loaded or traffic channel power < SSB power }}\textbf{: } A linear spectral extrapolation from SSB will result in an EMF that is higher than the real value.
	\item \textit{\textbf{Cell is fully loaded and/or traffic channel power > SSB power }}\textbf{: } A linear spectral extrapolation from SSB will result in lower than the real value.
\end{enumerate}

\subsection{Apriori Knowledge of Antenna Beam Pattern}
\begin{figure}[h]
	\centering
	\includegraphics[width=\linewidth]{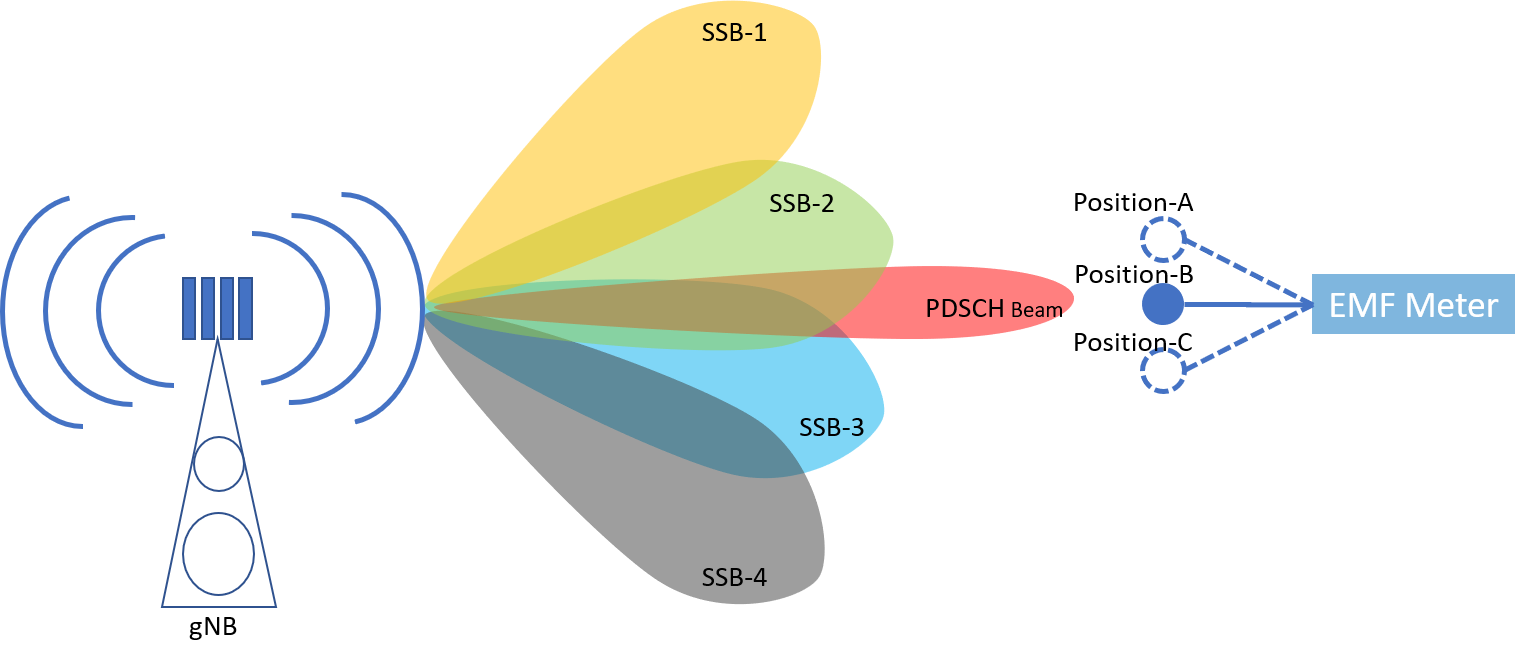}
	\caption{Beam sweeping of SSBs in a gNB and user-specific beamforming. It is important to have apriori information about the gNB antenna pattern for placing the isotropic antenna at an appropriate location in order to minimize the EMF measurement error. While measuring the EMF caused by traffic channel (user-specific beam) Position-B is desirable to get the accurate EMF.}
	\label{fig:ssb-beam-sweep-pdsch-beam}
\end{figure}

\begin{figure}[h]
	\centering
	\includegraphics[width=\linewidth]{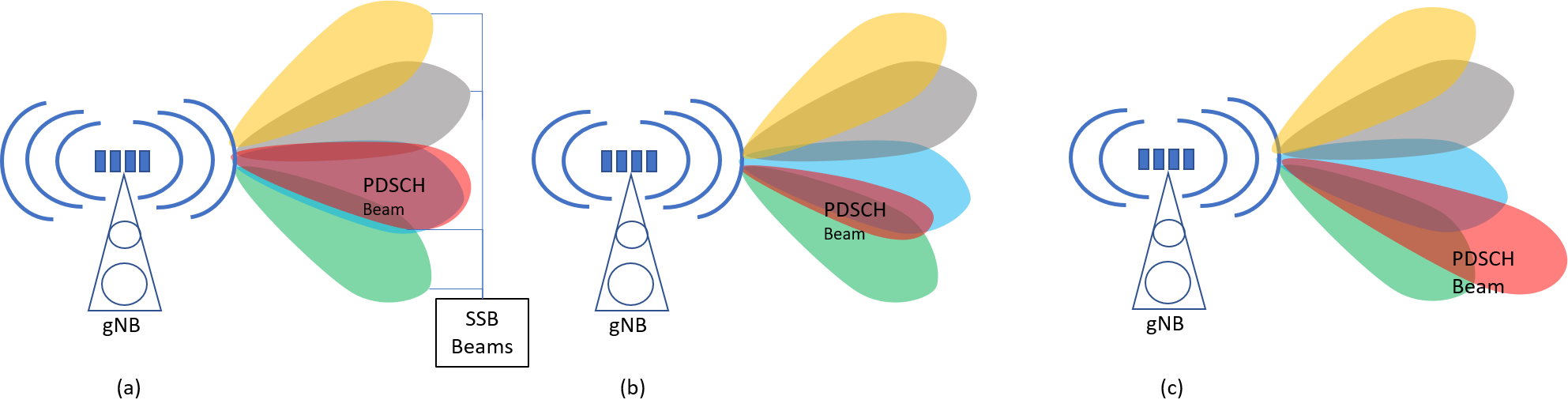}
	\caption{(a)Power and direction of traffic and SSB beam are same -- linear extrapolation on the power of SSB beam provides correct result (b),(c) Power and direction of traffic beam are different than SSB beam -- linear extrapolation on the power of SSB beam may provide incorrect result.}
	\label{fig:ssb-traffic-beam-power-misalignment}
\end{figure}

Once the process of beam sweeping is completed and traffic flow is started between the gNB and UE, a UE-specific beam is created by the gNB towards the UE. The width of this beam is narrower and the power content of this beam is higher compared to the SSB beam. To measure the EMF content in this traffic beam, it is important to have apriori information about the antenna pattern and maximum transmit power. Such information helps for the proper placement of the isotropic EMF measurement antenna during EMF measurement from the user-specific beam. Ideally, the isotropic should be placed at the center of the beam. Such information could be difficult to obtain in practice. A pictorial description of the same is shown in Figure-\ref{fig:ssb-beam-sweep-pdsch-beam}. 
Another challenge that can be encountered in practice is the directional and power mismatch between the beam used for SSB and the beam used for traffic channel(PDSCH and PDCCH). Dynamicity of the UE-specific beam (when the environment is changing, UE is moving, etc) makes the measurement further complicated in this situation. Three different cases may arise (a pictorial representation is shown in Figure-\ref{fig:ssb-traffic-beam-power-misalignment}). 

\begin{enumerate}[label=(\alph*)]
	\item \textit{\textbf{Traffic beam has the same power and same direction as SSB beam}}\textbf{: } In this case the linear extrapolation of the EMF measured from SSB provides a realistic EMF value of the traffic beam. 
	\item \textit{\textbf{Traffic beam has less power and direction not aligned with the SSB beam}}\textbf{: } In this case the linear extrapolation of the EMF measured from SSB will result in a higher than actual EMF value of traffic beam.
	\item \textit{\textbf{Traffic beam has more power and direction not aligned with the SSB beam}}\textbf{: } In this case the linear extrapolation of the EMF measured from SSB will result in a lower than actual EMF value of traffic beam.
\end{enumerate}

\subsection{Uplink-Downlink Relation Factor}
\begin{figure}[h]
	\centering
	\includegraphics[width=\linewidth]{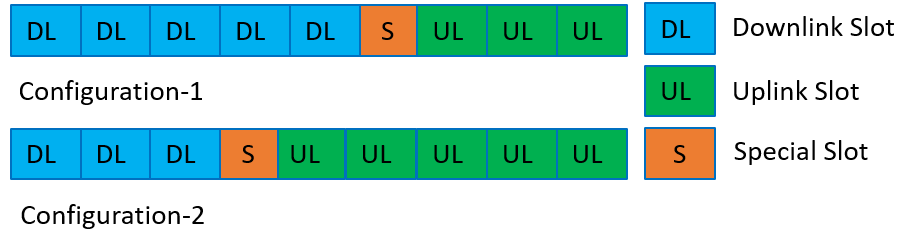}
	\caption{In the TDD mode of operation, it is important to perform a gated measurement to measure DL specific EMF.}
	\label{fig:uplink-downlink-relation-factor}
\end{figure}

TDD mode of operation is dominant in the current 5G deployment. The configuration of Downlink (DL) and Uplink (UL) slots could be different depending on which configuration has been agreed on by the operators. While measuring the EMF from the transmission coming from gNB, it is important to know the periodicity of DL and UL slots. Without such knowledge, the probe may log the cumulative EMF of both the DL and UL transmissions. A gated measurement could be required which triggers to measure of the power only when there is DL transmission. A pictorial depiction is shown in Figure-\ref{fig:uplink-downlink-relation-factor}. This is different from FDD where the probe can be simply tuned to the DL frequency for DL EMF measurement. 

\section{Field Experiments and Observations}\label{sec:field-exp}
\subsection{Field Experiments}
Concerning the radiation caused by gNB, we are interested in the radiated far-field region of the active antenna and for that either the measurement of electric field strength (E, unit V/m$^{2}$) or magnetic field strength (H, unit A/m) is sufficient. Optionally power density (unit W/m$^{2}$) is also used to quantify the radiation. We have chosen to work with electric field strength due to simplicity in measurement with the instruments we have, nonetheless, power density can also be used. To measure the EMF values, we have used Wave control SMP2 which is a broadband EMF measurement meter. This comprised an electric field isotropic probe, connected to the main control unit which records the electric field values for the band encompassing DC to 60 GHz. 

In this section, we will discuss the EMF measurement experiments which we performed to validate some of the discussions which we did in section-\ref{sec:5g-emf-concerns} and section-\ref{sec:5g-emf-meas-chan}. We did several experiments to measure the EMF at various distances from the UE during both DL and UL transmission. In all the experiments we perform a 10-second moving average for the entire duration of the measurement (for ruling out the outliers) and log the maximum and RMS value of the EMF in V/m$^{2}$. 
\subsection{Downlink EMF Measurement at UE from beamformed traffic}\label{subsec-downlink-emf-meas}
\subsubsection{Experiment Objective} In this experiment our objective is to measure and analyze the worst case (a) \textit{EMF observed at the UE} and (b) \textit{EMF observed in the vicinity of the UE} caused by DL waveform transmitted from a gNB which is equipped with an AAS. For both the EMF measurements (a) and (b) we compare two scenarios where in the first one, the gNB is equipped with an AAS, i.e., capable of beamforming, while in the second case, the gNB will not use an AAS. In both scenarios, the UE is not capable of performing any beamforming. Next, we detail the experimental setup.  

\subsubsection{Experiment Set-up}\label{subsubsec:exp-set-up}
\begin{figure}[h]
	\centering
	\includegraphics[width=\linewidth]{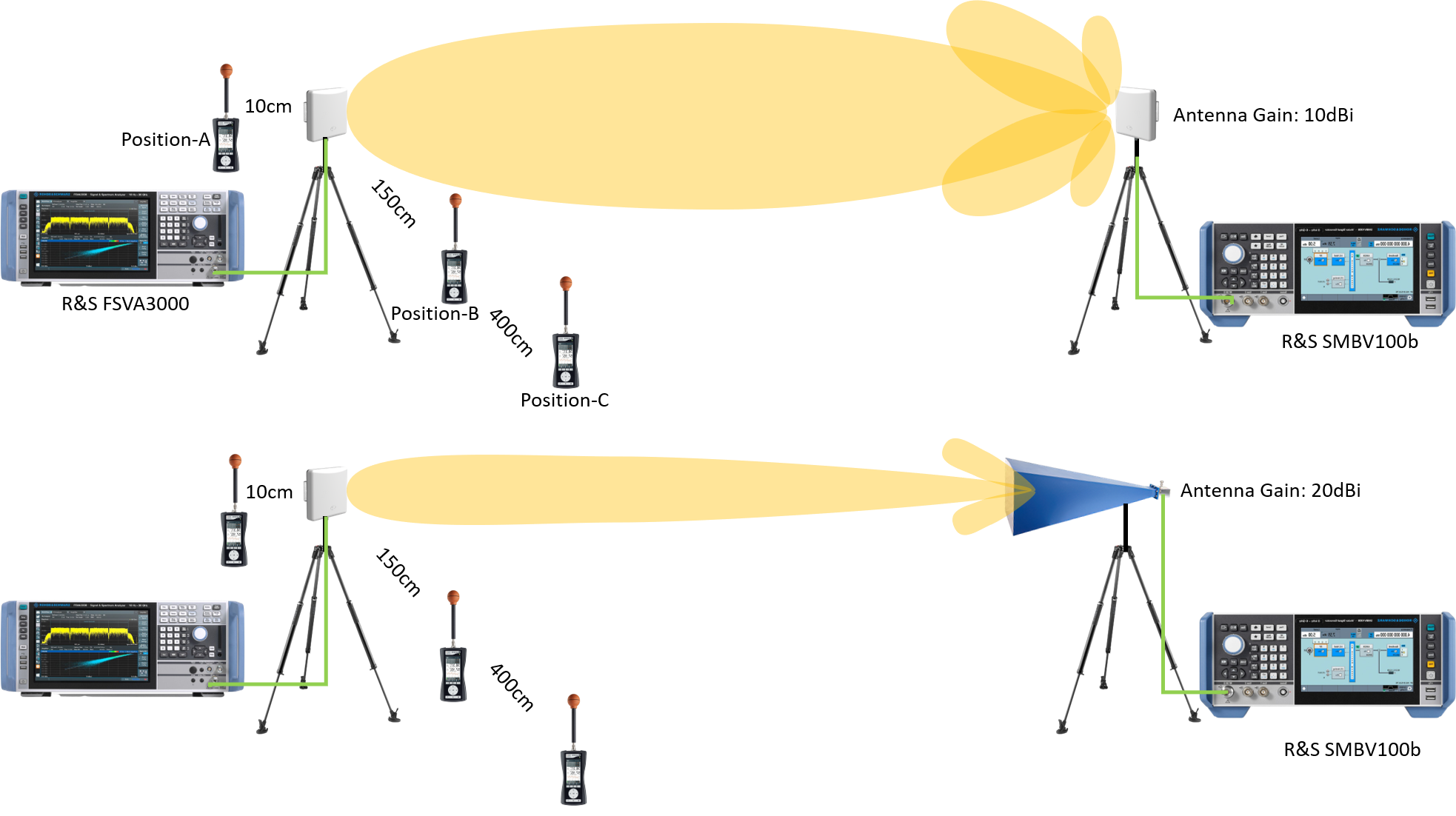}
	\caption{Experiment set-up for downlink EMF measurement as discussed in section-\ref{subsubsec:exp-set-up}}
	\label{fig:beam-vs-patch}
\end{figure}

For these experiments, we have used emulated gNB and emulated UE. This is a unidirectional connection where only the gNB makes a transmission beam towards the UE while the UE simply decodes the transmission from UE. To emulate the gNB we have used a commercial 5G vector signal generator from Rhode \& Schwartz (SMBV100B) which is capable of generating standard-compliant 5G-NR TDD waveforms. Our emulated gNB using a commercial 5G signal generator cannot perform beam sweeping, i.e., multiple SSB in different directions. Hence, to emulate an active-antenna gNB system that creates a dedicated beam towards the UE for traffic and data channels, we used a high directive horn antenna at the NBA. This is not an accurate representation, however, it closely emulates the situation of beam correspondence. Beam-correspondence is the step where a user-specific traffic beam is created from the gNB and the actual downlink traffic proceeds. To compare, we also used a low directive antenna at gNB which emulates conventional wide-angled transmission from the gNB. Next, to emulate the UE, we have used a commercial 5G signal analyzer from Rhode \& Schwartz (FSV3000) capable of decoding the test waveforms from the gNB. On the UE side, we are simply using an antenna with low directivity which is a valid assumption. The transmission frequency for our experiment is $ 3.775 $ GHz for which we have the license from Institut Luxembourgeois de Régulation(ILR) for experimental work. To measure the EMF, we have used a broadband-EMF meter and an isotropic probe. Our measurement was dedicated to a single transmitter (gNB) which was under our control, hence, we decided to use broadband measurement. The probe was mounted on a wooden tripod at a height of $1.5$m and recordings were time-averaged for 6 minutes. Although the latest ICNIRP guideline has a stated duration of 30 minutes for the measurement of EMF\cite{international2020guidelines}, in our case (a) the gNB is fully under our control (b) there is only one UE that is served by the gNB (c) the environment is static and no other 5G transmitters nearby. Hence, we decided to use 6 minutes of dwell time for our measurements. A pictorial representation of the emulated gNB and UE is shown in \ref{fig:beam-vs-patch} and Table-\ref{tab:2}. Next, we will discuss the experimental methodology. 
\subsubsection{Experiment Methodology}\label{subsubsection:experiment-methodology}
\begin{itemize}
	\item \textbf{Step-1:} We connect the gNB with a low-directive antenna and connect the UE with a similar low-directive antenna 
	\item \textbf{Step-2:} A known waveform is transmitted with a fixed power TxP from the gNB towards UE and the PDSCH Error Vector Magnitude(EVM) is recorded for that TxP. We start with a transmission bandwidth of 5 MHz.   
	\item \textbf{Step-3:} For the known TxP and EVM, the EMF is measured (\textit{Maximum} and \textit{RMS} value) at three different locations: 
\begin{enumerate}[label=(\alph*)]
	\item $ 10 $cm near to the UE: This represents the EMF observed by a user which is very near to the human body (for example head).  
	\item $ 150 $cm away from UE: This represents the EMF observed by a user who is in the vicinity of the human body (for example a person talking on speakerphone). 
	\item $ 400 $cm away from the UE: This represents the EMF observed by a UE which is far away from the human body (for example a person using a smartphone as a wifi hotspot). 
\end{enumerate}
	\item \textbf{Step-4:} We connect the high-directive horn antenna to the gNB and keep the rest of the set-up intact. 
	\item \textbf{Step-5:} Now we perform the again steps-2 and step-3 and log the EMF values at different locations as mentioned in step-4. 
	\item \textbf{Step-6:} We repeat step-2, 3, 4 and 5 for $ 10 $, $ 20 $ and $ 50 $ MHz of transmission bandwidth. 
\end{itemize}
\begin{table}[h]
    \centering 
	\caption{ Details of equipment and experiment set-up as discussed in section-\ref{subsubsection:experiment-methodology}}
	\label{tab:2}
	\begin{tabular}{|c|p{5.4cm}|}
		\hline
		\textbf{Parameters}  & \textbf{Values}         \\ \hline
		gNB                  & R\&S SMBV100b      \\ \hline
		nrUE                 & R\&S FSV3000          \\ \hline
		Frequency & 3775 MHz  \\ \hline
		Bandwidth            & 5, 10, 20, 50 MHz                  \\ \hline
		gNB Antenna          & WR-229, 3.3-4.9 GHz, 20dBi, Horizontal HPBW $16.7^\circ$, Vertical HPBW $17.1^\circ$ \\ \hline
		nrUE Antenna         & W36-CP-9, 3.4-3.8 GHz, 10dBi, Horizontal HPBW $50^\circ$, Vertical HPBW $50^\circ$    \\ \hline
		Duplex Mode          & TDD                     \\ \hline
		Broadband EMF meter           & Wavecontrol SMP2 DC-60GHz, WPF-40 Field Probe 1MHz - 40GHz                           \\ \hline		
	\end{tabular}
\end{table}

\begin{table*}[]
	\caption{Downlink EMF measurement for experiments in section-\ref{subsec-downlink-emf-meas}}
	\label{tab:3}
	\centering
	\begin{tabular}{|p{0.5cm}|p{4.4cm}|p{2.4cm}|p{1cm}|p{1.4cm}|p{1.1cm}|p{1.1cm}|p{1.1cm}|}
		\hline
		&
		\cellcolor[HTML]{FFFFFF}\textbf{Transmission Mode} &
		\cellcolor[HTML]{FFFFFF}\textbf{gNB Antenna Type} &
		\cellcolor[HTML]{FFFFFF}\textbf{TxP (dBm)} &
		\cellcolor[HTML]{FFFFFF}\textbf{Distance of probe from UE (cm)} &
		\cellcolor[HTML]{FFFFFF}\textbf{PDSCH EVM (\%)} &
		\cellcolor[HTML]{FFFFFF}\textbf{Max-EMF (V/m)} &
		\cellcolor[HTML]{FFFFFF}\textbf{RMS-EMF (V/m)} \\ 
         \hline 
	  \rowcolor[HTML]{FFFFFF}	R1.1 &
		  \cellcolor[HTML]{FFFFFF} &
	    \cellcolor[HTML]{FFFFFF} Low-directive &
		\cellcolor[HTML]{FFFFFF}14.7 &
		\cellcolor[HTML]{FFFFFF}10 &
		\cellcolor[HTML]{FFFFFF}3.75 &
		\cellcolor[HTML]{FFFFFF}1.64 &
		\cellcolor[HTML]{FFFFFF}1.214 \\
            \cline{1-1} \cline{3-8} 
		\rowcolor[HTML]{FFFFFF} R1.2 &
		\cellcolor[HTML]{FFFFFF} &
		\cellcolor[HTML]{FFFFFF} Low-directive &
		\cellcolor[HTML]{FFFFFF}14.7 &
		\cellcolor[HTML]{FFFFFF}150 &
		\cellcolor[HTML]{FFFFFF}3.75 &
		\cellcolor[HTML]{FFFFFF} 1.69 &
		\cellcolor[HTML]{FFFFFF}1.256 \\ \cline{1-1} \cline{3-8} 
		\rowcolor[HTML]{FFFFFF}  R1.3 &
		\multirow{-3}{*}{\cellcolor[HTML]{FFFFFF}Bandwidth 5MHz, SCS 15KHz, TDD} &
		\cellcolor[HTML]{FFFFFF} Low-directive &
		\cellcolor[HTML]{FFFFFF}14.7 &
		\cellcolor[HTML]{FFFFFF}400 &
		\cellcolor[HTML]{FFFFFF}3.75 &
		\cellcolor[HTML]{FFFFFF}1.62 &
		\cellcolor[HTML]{FFFFFF}1.272 \\
          \hline
		R2.1 &
		\cellcolor[HTML]{FFFFFF} &
		\cellcolor[HTML]{FFFFFF}Low-directive &
		\cellcolor[HTML]{FFFFFF}14.7 &
		\cellcolor[HTML]{FFFFFF}10 &
		\cellcolor[HTML]{FFFFFF}3.75 &
		\cellcolor[HTML]{FFFFFF}1.64 &
		\cellcolor[HTML]{FFFFFF}1.322 \\ \cline{1-1} \cline{3-8} 
		R2.2 &
		\cellcolor[HTML]{FFFFFF} &
		\cellcolor[HTML]{FFFFFF}Low-directive &
		\cellcolor[HTML]{FFFFFF}14.7 &
		\cellcolor[HTML]{FFFFFF}150 &
		\cellcolor[HTML]{FFFFFF}3.75 &
		\cellcolor[HTML]{FFFFFF}2.16 &
		\cellcolor[HTML]{FFFFFF}1.366 \\ \cline{1-1} \cline{3-8} 
		R2.3 &
		\multirow{-3}{*}{\cellcolor[HTML]{FFFFFF}Bandwidth 10MHz, SCS 15KHz,   TDD} &
		\cellcolor[HTML]{FFFFFF}Low-directive &
		\cellcolor[HTML]{FFFFFF}14.7 &
		\cellcolor[HTML]{FFFFFF}400 &
		\cellcolor[HTML]{FFFFFF}3.75 &
		\cellcolor[HTML]{FFFFFF}2.17 &
		\cellcolor[HTML]{FFFFFF}1.31 \\ \hline
		\rowcolor[HTML]{FFFFFF} 
		R3.1 &
		\cellcolor[HTML]{FFFFFF} &
		Low-directive &
		14.7 &
		10 &
		4.5 &
		2.28 &
		1.803 \\ \cline{1-1} \cline{3-8} 
		\rowcolor[HTML]{FFFFFF} 
		R3.2 &
		\cellcolor[HTML]{FFFFFF} &
		Low-directive &
		14.7 &
		150 &
		4.5 &
		2.31 &
		1.798 \\ \cline{1-1} \cline{3-8} 
		\rowcolor[HTML]{FFFFFF} 
		R3.3 &
		\multirow{-3}{*}{\cellcolor[HTML]{FFFFFF}Bandwidth 20MHz, SCS 15KHz,   TDD} &
		Low-directive &
		14.7 &
		400 &
		4.5 &
		2.27 &
		1.795 \\ \hline
		R4.1 &
		\cellcolor[HTML]{FFFFFF} &
		\cellcolor[HTML]{FFFFFF}Low-directive &
		\cellcolor[HTML]{FFFFFF}14.7 &
		\cellcolor[HTML]{FFFFFF}10 &
		\cellcolor[HTML]{FFFFFF}6.3 &
		\cellcolor[HTML]{FFFFFF}2.45 &
		\cellcolor[HTML]{FFFFFF}2.11 \\ \cline{1-1} \cline{3-8} 
		R4.2 &
		\cellcolor[HTML]{FFFFFF} &
		\cellcolor[HTML]{FFFFFF}Low-directive &
		\cellcolor[HTML]{FFFFFF}14.7 &
		\cellcolor[HTML]{FFFFFF}150 &
		\cellcolor[HTML]{FFFFFF}6.3 &
		\cellcolor[HTML]{FFFFFF}2.39 &
		\cellcolor[HTML]{FFFFFF}2.05 \\ \cline{1-1} \cline{3-8} 
		R4.3 &
		\multirow{-3}{*}{\cellcolor[HTML]{FFFFFF}Bandwidth 50MHz, SCS 15KHz,   TDD} &
		\cellcolor[HTML]{FFFFFF}Low-directive &
		\cellcolor[HTML]{FFFFFF}14.7 &
		\cellcolor[HTML]{FFFFFF}400 &
		\cellcolor[HTML]{FFFFFF}6.3 &
		\cellcolor[HTML]{FFFFFF}2.41 &
		\cellcolor[HTML]{FFFFFF}2.01 \\ \hline
		\rowcolor[HTML]{FFFFFF} 
		R5.1 &
		\cellcolor[HTML]{FFFFFF} &
		High-directive Horn &
		9.8 &
		10 &
		0.42 &
		1.58 &
		1.114 \\ \cline{1-1} \cline{3-8} 
		\rowcolor[HTML]{FFFFFF} 
		R5.2 &
		\cellcolor[HTML]{FFFFFF} &
		High-directive Horn &
		9.8 &
		150 &
		0.42 &
		1.01 &
		0.93 \\ \cline{1-1} \cline{3-8} 
		\rowcolor[HTML]{FFFFFF} 
		R5.3 &
		\multirow{-3}{*}{\cellcolor[HTML]{FFFFFF}Bandwidth 5MHz, SCS 15KHz, TDD} &
		High-directive Horn &
		9.8 &
		400 &
		0.42 &
		0.81 &
		0.73 \\ \hline
		R6.1 &
		\cellcolor[HTML]{FFFFFF} &
		\cellcolor[HTML]{FFFFFF}High-directive Horn &
		\cellcolor[HTML]{FFFFFF}9.8 &
		\cellcolor[HTML]{FFFFFF}10 &
		\cellcolor[HTML]{FFFFFF}0.5 &
		\cellcolor[HTML]{FFFFFF}2.1 &
		\cellcolor[HTML]{FFFFFF}1.21 \\ \cline{1-1} \cline{3-8} 
		R6.2 &
		\cellcolor[HTML]{FFFFFF} &
		\cellcolor[HTML]{FFFFFF}High-directive Horn &
		\cellcolor[HTML]{FFFFFF}9.8 &
		\cellcolor[HTML]{FFFFFF}150 &
		\cellcolor[HTML]{FFFFFF}0.5 &
		\cellcolor[HTML]{FFFFFF}1.27 &
		\cellcolor[HTML]{FFFFFF}0.95 \\ \cline{1-1} \cline{3-8} 
		R6.3 &
		\multirow{-3}{*}{\cellcolor[HTML]{FFFFFF}Bandwidth 10MHz, SCS 15KHz,   TDD} &
		\cellcolor[HTML]{FFFFFF}High-directive Horn &
		\cellcolor[HTML]{FFFFFF}9.8 &
		\cellcolor[HTML]{FFFFFF}400 &
		\cellcolor[HTML]{FFFFFF}0.5 &
		\cellcolor[HTML]{FFFFFF}1.16 &
		\cellcolor[HTML]{FFFFFF}0.76 \\ \hline
		\rowcolor[HTML]{FFFFFF} 
		R7.1 &
		\cellcolor[HTML]{FFFFFF} &
		High-directive Horn &
		9.8 &
		10 &
		0.63 &
		2.15 &
		1.6 \\ \cline{1-1} \cline{3-8} 
		\rowcolor[HTML]{FFFFFF} 
		R7.2 &
		\cellcolor[HTML]{FFFFFF} &
		High-directive Horn &
		9.8 &
		150 &
		0.63 &
		1.33 &
		1.31 \\ \cline{1-1} \cline{3-8} 
		\rowcolor[HTML]{FFFFFF} 
		R7.3 &
		\multirow{-3}{*}{\cellcolor[HTML]{FFFFFF}Bandwidth 20MHz, SCS 15KHz,   TDD} &
		High-directive Horn &
		9.8 &
		400 &
		0.63 &
		1.1 &
		1.05 \\ \hline
		R8.1 &
		\cellcolor[HTML]{FFFFFF} &
		\cellcolor[HTML]{FFFFFF}High-directive Horn &
		\cellcolor[HTML]{FFFFFF}9.9 &
		\cellcolor[HTML]{FFFFFF}10 &
		\cellcolor[HTML]{FFFFFF}1.26 &
		\cellcolor[HTML]{FFFFFF}2.33 &
		\cellcolor[HTML]{FFFFFF}1.85 \\ \cline{1-1} \cline{3-8} 
		R8.2 &
		\cellcolor[HTML]{FFFFFF} &
		\cellcolor[HTML]{FFFFFF}High-directive Horn &
		\cellcolor[HTML]{FFFFFF}9.9 &
		\cellcolor[HTML]{FFFFFF}150 &
		\cellcolor[HTML]{FFFFFF}1.26 &
		\cellcolor[HTML]{FFFFFF}1.42 &
		\cellcolor[HTML]{FFFFFF}1.4 \\ \cline{1-1} \cline{3-8} 
		R8.3 &
		\multirow{-3}{*}{\cellcolor[HTML]{FFFFFF}Bandwidth 50MHz, SCS 15KHz,   TDD} &
		\cellcolor[HTML]{FFFFFF}High-directive Horn &
		\cellcolor[HTML]{FFFFFF}9.9 &
		\cellcolor[HTML]{FFFFFF}400 &
		\cellcolor[HTML]{FFFFFF}1.26 &
		\cellcolor[HTML]{FFFFFF}1.17 &
		\cellcolor[HTML]{FFFFFF}1.1 \\ \hline
	\end{tabular}
\end{table*}

\subsubsection{Observations}
The measurements are listed in Table-\ref{tab:3} and from that we observe the following:
\begin{enumerate}
	\item For a given transmit power TxP:
	\begin{enumerate}
		\item For the low-directive antenna, the EMF is almost the same for the three locations a, b and c. This is observed for all the transmission modes.
		\item For the high-directive horn antenna, the EMF decreases as the probe is moved from a$ \rightarrow $b$ \rightarrow $c. This is also observed for all the transmission modes.   
	\end{enumerate}
	\item The behavior above is consistent with respect to the increase of the transmission bandwidth. As the transmission bandwidth is increased from 5$ \rightarrow $10$ \rightarrow $20$ \rightarrow $50 MHz, the EMF increases, however, it decreases as the probe is moved away from the UE. 
	\item Using high-directive horn antenna at the gNB, lower EVM (compared to the low-directive antenna at the gNB) is achieved at the UE even with a lower transmit power TxP. For example: In R1.1, 1.2, and 1.3 the EVM is 3.75 at a TxP of 14.7 dBm, while for its counterparts in R5.1, 5.2, and 5.3 the EVM is 0.42 at TxP of 9.8 dBm. 
	\item From 1(a) and 1(b), it can be said that when a gNB makes a UE-specific beam, the EMF in the vicinity of the UE is reduced. 
	\item Further, from 2, it can be said that the SNR of the signal can be improved with UE-specific beamforming and at the same time, the EMF in the vicinity of the UE is decreased. 
\end{enumerate}

\subsection{Uplink EMF}\label{subsec-uplink-emf-meas}
\subsubsection{Experiment Objective}
In the section-\ref{subsec-high-uplink-traffic}, we discussed how the traffic pattern has changed, and now the uplink traffic is comparable to downlink traffic. The objective of this experiment is to verify the fact that UE may experience very high EMF during audio and video calls when the RSRP is poor. We do not focus on a particular operator, hence, we use the broadband meter for the uplink measurement. 

\subsubsection{Experiment Set-up}
\begin{figure}[]
	\centering
	\includegraphics[width=\linewidth]{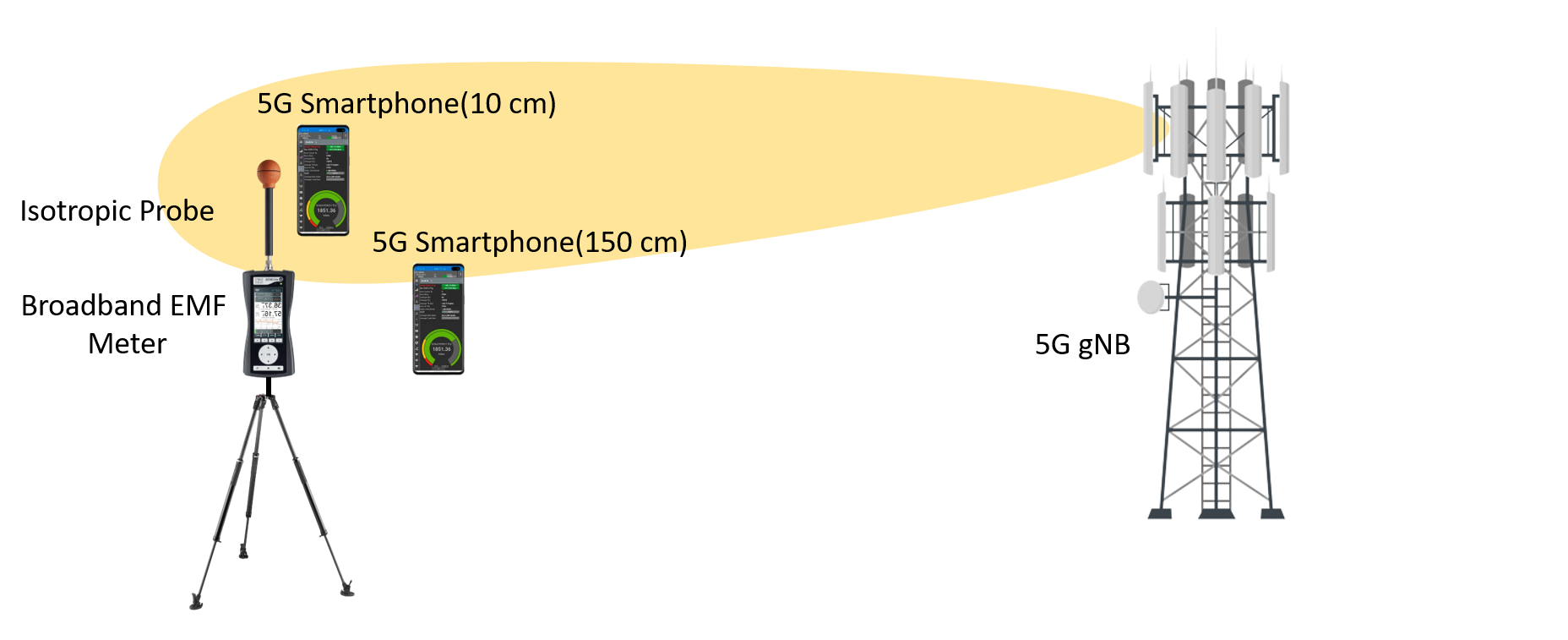}
	\caption{Uplink EMF measurement very near to the UE and its vicinity}
	\label{fig:uplink-meas}
\end{figure}

For this experiment, we have used a COTS smartphone as our UE which is connected to a 5G base station. During the uplink traffic between COTS UE and the gNB, the isotropic probe is placed at a height of 1.5 meters from the ground on a tripod. A pictorial representation is shown in Figure-\ref{fig:uplink-meas}. For measuring the RSRP in the COTS UE, we have used Qualipoc\cite{rhode-sch-qualipoc}.   
\subsubsection{Experiment Methodology}
The experiment was conducted in the following steps:
\begin{itemize}
	\item \textbf{Step-1:} The UE was connected to a gNB under low RSRP conditions. \textit{For our in-lab measurement, the UE showed an RSRP fluctuating between -111 dBm to -100 dBm as shown by Qualipoc.}
	\item \textbf{Step-2:} The isotropic probe was placed at a distance of 10 cm from the UE.
	\item \textbf{Step-3:} Traffic was initiated between UE and gNB through audio and video calls. The EMF meter was turned on to measure 6 mins. 
	\item \textbf{Step-4:} Step-2 was repeated with the probe placed at a distance of 150 cm from the UE. 
	\item \textbf{Step-5:} The set-up was relocated to an area with better RSRP and steps-1,2,3,4 were repeated. 
\end{itemize}
In all the experiments above, time averaging was performed to eliminate the floating traffic error in 5G, i.e., assuming that the cell is loaded with traffic. Further, each mobile phone is equipped with an Automatic Power Control system that adjusts the output power level to the minimum required value necessary to establish a connection with a base station. Besides, the UE goes thru DRX and TDD (duty cycle), hence, time averaging was necessary.  
\begin{table*}[h]
	\caption{Uplink EMF measurement for experiments in section-\ref{subsec-uplink-emf-meas}}
	\label{tab:4}
	\centering
	\begin{tabular}{l|c|cc|cc|}
		\cline{2-6}
		\textbf{} &
		\textbf{RSRP (dBm)} &
		\multicolumn{2}{c|}{-111} &
		\multicolumn{2}{c|}{-96} \\ \cline{2-6} 
		\textbf{} &
		\textbf{Distance of probe from UE (cm)} &
		\multicolumn{1}{c|}{10} &
		150 &
		\multicolumn{1}{c|}{10} &
		150 \\ \hline
		\multicolumn{1}{|l|}{\textbf{Case-1}} &
		\textbf{Max / RMS EMF at no activity (V/m)} &
		\multicolumn{1}{c|}{11.57 / 5.547} &
		3.32/1.692 &
		\multicolumn{1}{c|}{5.62 / 2.447} &
		1.75/0.991 \\ \hline
		\multicolumn{1}{|l|}{\textbf{Case-2}} &
		\textbf{Max / RMS EMF at audio traffic (V/m)} &
		\multicolumn{1}{c|}{19.44 / 5.899} &
		3.06/1.676 &
		\multicolumn{1}{c|}{6.99 / 2.518} &
		1.88/1.011 \\ \hline
		\multicolumn{1}{|l|}{\textbf{Case-3}} &
		\textbf{Max / RMS EMF at video traffic (V/m)} &
		\multicolumn{1}{c|}{21.22 / 11.33} &
		3.98/1.881 &
		\multicolumn{1}{c|}{7.12 / 3.162} &
		2.01/1.011 \\ \hline
	\end{tabular}
\end{table*}

\begin{figure}[h]
	\centering
	\includegraphics[width=0.9\linewidth]{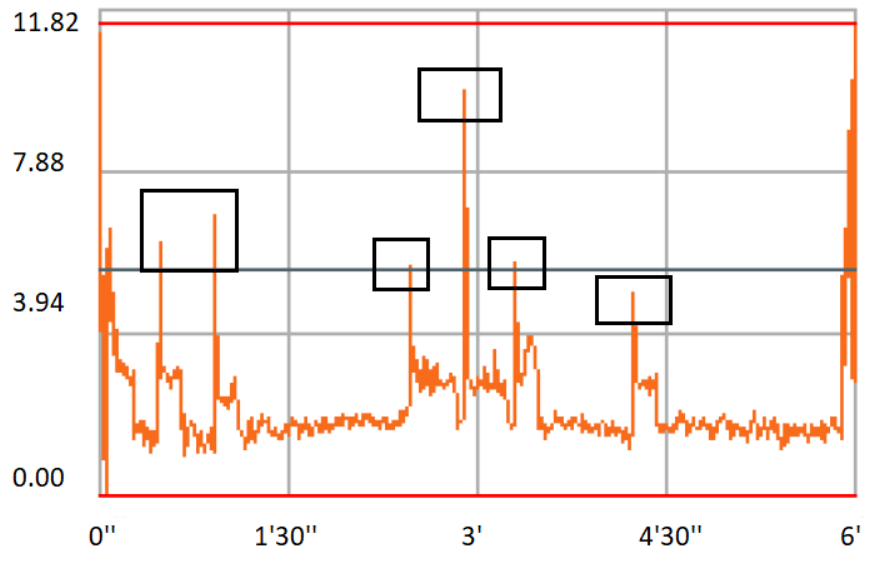}
	\caption{Screenshot of uplink EMF measurement from Wavecontrol-SMP2 during the audio call. The X-axis represents time while Y-axis represents EMF(V/m). Spikes in the EMF measurement can be seen as the user speaks while keeping the UE near to the head.}
	\label{fig:voice-call-spike}
\end{figure}

\begin{figure}[h]
	\centering
	\includegraphics[width=0.9\linewidth]{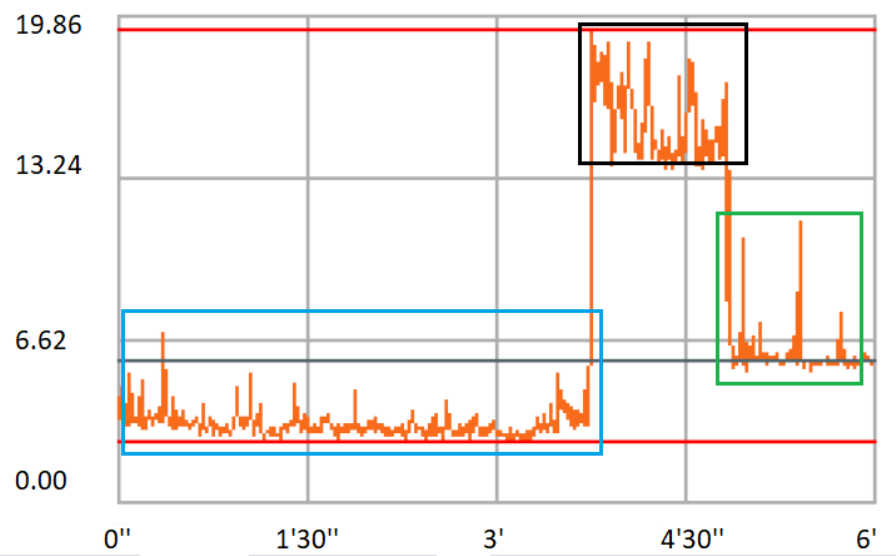}
	\caption{Screenshot of uplink EMF measurement from Wavecontrol-SMP2 during a video call. The X-axis represents time while Y-axis represents EMF(V/m). Spikes in the EMF measurement can be seen as the user makes movement neat the UE camera while keeping the UE near to head. The area enclosed in the black rectangle represents an elevated EMF level when the user makes activity near the UE camera, the area enclosed in the green rectangle represents an elevated EMF level when a user at makes activity near the gNB camera, and the area enclosed in blue rectangle represents EMF level when none of the users make any activity near their respective cameras. }
	\label{fig:video-call-spike}
\end{figure}

\subsubsection{Observations}
The EMF measurement from the experiments is listed in Table-\ref{tab:4}. We observe the following:
\begin{enumerate}
	\item When the RSRP is weak at the UE (due to a large distance from the gNB or the user is inside a building or a car), the UE transmits at a higher power which leads to a high EMF experience by the user. This behavior is consistent in all three cases. Even during no activity, there is an exchange of messages between the UE and gNB, which leads to excessive EMF measurement during case-1, i.e., No Activity. Maximum and RMS EMF in such cases exceed the defined limit of 3V/m \cite{vdl-emf}. 
	\item Probe when placed near to UE (10 cm) experiences higher EMF when placed far (150 cm) from the UE. This behavior is also consistent in all three cases. Thus it can be said that using the smartphone away from the body is relatively safe. 
	\item Furthermore, two interesting phenomena were observed while placing audio and video calls: 
	\begin{itemize}
		\item When the user at the UE speaks during an audio call, there are sudden jumps in the EMF which is shown in Figure-\ref {fig:voice-call-spike}. This can be explained by the fact that when the user speaks, uplink traffic is generated. It is also observed that EMF does not exceed when there is no voice activity from the UE side. This behavior is observed only when the UE is placed near the probe. At a distance far from the probe, such spikes are not observed.   
		\item When the user at the UE makes an activity during a video call, there are sudden jumps in the EMF which is shown in a black rectangle in Figure-\ref {fig:video-call-spike}. Similarly, when the person at the gNB side makes activity, the EMF level elevates as shown in a green rectangle in Figure-\ref {fig:video-call-spike}. These can be explained by the fact that when the user on the UE side makes activity near the smartphone camera, new video frames are generated for uplink traffic and sent, and vice versa for the gNB side. It is also observed that EMF does not exceed when there is no video activity on both sides, i.e., a video call in progress but none of the users are facing the camera. This is shown in a blue rectangle in Figure-\ref{fig:video-call-spike}. This behavior is observed only when the UE is placed near the probe. At a distance far from the probe, such elevated levels of EMF are not observed.  
	\end{itemize}
\end{enumerate}


\section{EMF Measurements under spatio-temporal variations}\label{sec:open-challenges}
Spatio-temporal characteristics are important factors to be considered while measuring 5G EMF which exclusively uses AAS and dynamic-TDD. Such characteristics must be considered while performing the measurement of EMF and post-processing. This aspect can be motivated by the following two aspects: 
\begin{itemize}
	\item Application of higher frequency bands such as mmWave and terahertz (THz) require beamforming techniques to cope with the higher path losses (Section-\ref{sec:5g-emf-concerns}). Besides, beam-tracking and beam-steering are also needed for mobile users and this is achieved using AAS at the gNB. mmWave channel is mainly dominated by a limited number of paths from \emph{different angle of arrivals (AoAs)}\cite{samimi20163,sheemar2021hybrid}, as shown in Figure \ref{AoAs}. Thus, different directions may experience different EMF radiation, i.e., directional dependency of the EMF. Moreover, as the user moves, the directional dependency also varies. However, the existing EMF assessment techniques (Frequency selective, Code Selective, and Broadband \cite{emf-meas-methd}), do not account for the directional dependency. This motivates the development of direction/AoAs dependent EMF exposure assessment techniques (and devices) in which the spatial characteristics of the EMF are also taken into account such as EMF probes with multi-antenna.  
	\begin{figure}
		\centering
		\includegraphics[width=7.5cm,height=5cm]{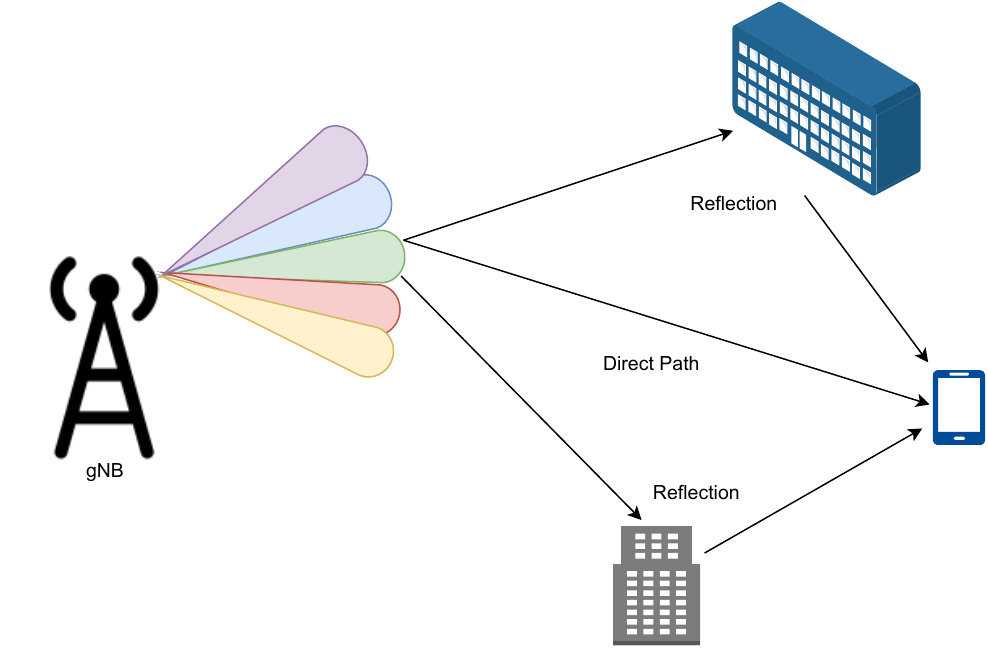}
		\caption{Propagation from gNB to the receiver with direct and reflected paths in mmWave.}
		\label{AoAs}
	\end{figure}
	
	\item Another critical aspect is Dynamic TDD (varying duty cycles for uplink and downlink) which is provisioned in 5G to deal with the highly variable data demands. This may lead to dynamic variations in the EMF exposure over time. To capture such dynamicity of EMF over time, long-duration EMF measurement is required which could be time-consuming and expensive. Thus, extrapolation techniques, which also take into account the temporal dynamicity of the duty cycle of 5G-NR transmissions, are required to obtain long-term data from the measurements done over a short period. 
\end{itemize}

\section{Conclusions and Ongoing Work}\label{sec:conclusions}

Careful planning and safe deployment of 5G technologies will bring enormous benefits to both society and the economy. Active Antenna Systems are the key enabler for 5G technologies such as Massive MIMO, mmWave, and Dynamic Beamforming. Unfortunately, superficial knowledge about these technologies has created concerns among the general public over EMF radiation. Several standardization bodies are studying the effects of EMF radiation on the human body and set limits on the transmission by the 5G base stations. However, over-conservative limits may hamper the 5G rollout and the capacity it can achieve. Using the conventional EMF measurement methods may lead to misleading conclusions that 5G frames are more dynamic and user-specific. Hence, it is imperative to study and find the gaps in conventional EMF measurement methodology and come up with adaptations/improvements. 

In this work, we have focused on the measurement of EMF caused by using an AAS at the gNB. We observed that an AAS, through UE-specific beamforming can achieve same performance as a sectoral-antenna (low directive), however, reducing the EMF radiation in the vicinity of the UE. Further, only the transmission from base stations towards the users, i.e., downlink, are being perceived as harmful. Through field experiments, we observed that uplink EMF radiation can exceed the limits by many orders especially when the RSRP at the UE is poor. Finally, we provide guidelines to develop novel EMF measurement methodology which also takes spatio-temporal characteristics of the 5G transmissions into account. This will be helpful for mmWave and dynamic TDD based operations.  
propose 

We believe that the study and field experiments conducted in this work will help in providing clarity over the EMF concerns raised by the general public. Nonetheless, the work is in progress and we have several planned activities some of which are: 
\begin{enumerate}
    \item UAV-based antenna pattern measurement of AAS: An apriori knowledge about the antenna pattern is critical for proper placement of the measurement probes. 
    \item EMF measurement with cell-barring at the gNB: Through cell barring, a dedicated UE-specific beam can be formed for a duration suitable enough to assess the worst-case EMF exposure.  
\end{enumerate}

\bibliographystyle{IEEEtran}
\bibliography{EMF}

\begin{IEEEbiography}[{\includegraphics[width=1.5in,height=1.25in,clip,keepaspectratio]{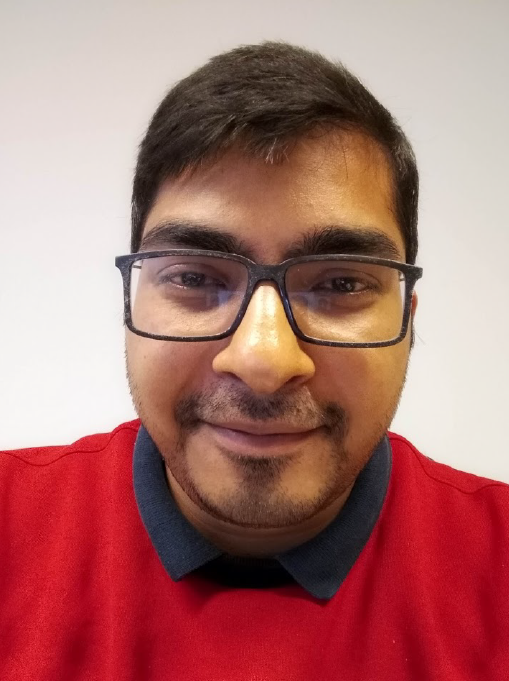}}]  {\textbf{Sumit Kumar}}  (S’14-M’19) received the
B.Tech and M.S.  in Electronics $\&$ Communication
Engineering from Gurukula Kangri University,
Haridwar, India (2008) and the International
Institute of Information Technology, Hyderabad,
India (2014), respectively, and the Ph.D. from Eurecom
(France) in 2019. Currently, he is working
as a Research Associate at the Interdisciplinary
Centre for Security, Reliability, and Trust (SnT),
University of Luxembourg. His research interests
are in wireless communication, interference management, Integration of 5G
with Non-Terrestrial-Networks and Software Defined Radio prototyping.
\end{IEEEbiography}

\begin{IEEEbiography}[{\includegraphics[width=1in,height=5in,clip,keepaspectratio]{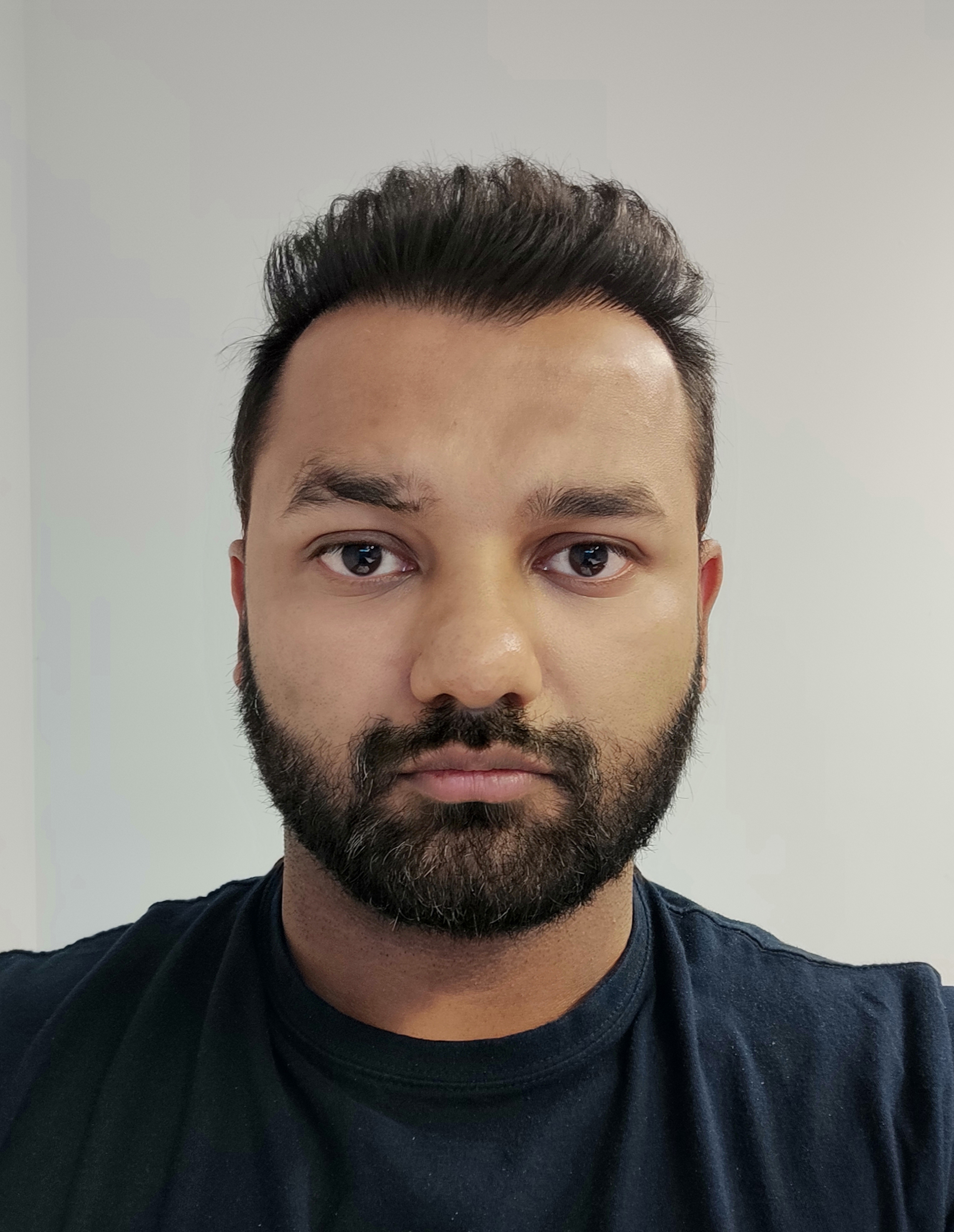}}]  {\textbf{Chandan Kumar Sheemar}} received his Bachelor in Information Engineering and Master in Telecommunications Engineering degrees in 2016 and 2018 from the University of Padua, Italy. He received his PhD in signal processing for wireless communications from EURECOM, France, in 2022, under the supervision of Prof. Dirk Slock. The topic of his PhD thesis was massive MIMO full duplex communications, for which he developed several novel hybrid beamforming designs to deal with the self-interference in the millimeter wave. He is currently a research associate at the University of Luxembourg in the Signal Processing $\&$ Satellite Communications research group, SIGCOM, headed by Prof. Symeon Chatzinotas. His research interests include Statistical Signal Processing, Optimization Theory, Beamforming, Intelligent Reflecting Surfaces and Non-Terrestrial Networks.
\end{IEEEbiography}

\begin{IEEEbiography}[{\includegraphics[width=1in,height=1.4in,clip,keepaspectratio]{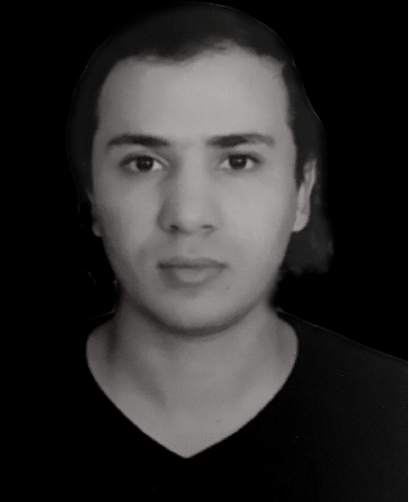}}]  {\textbf{ABDELRAHMAN ASTRO}}  received the B.Sc.
degree in electrical engineering from the Canadian
International College, Egypt, in 2017. He is currently
an RnD Specialist at the Interdisciplinary
Centre for Security, Reliability and Trust, University
of Luxembourg. His research interests include
software-defined radios (SDR), wireless communication
systems, real-time systems, and systems
architecture.
\end{IEEEbiography}

\begin{IEEEbiography}[{\includegraphics[width=1in,height=1.4in,clip,keepaspectratio]{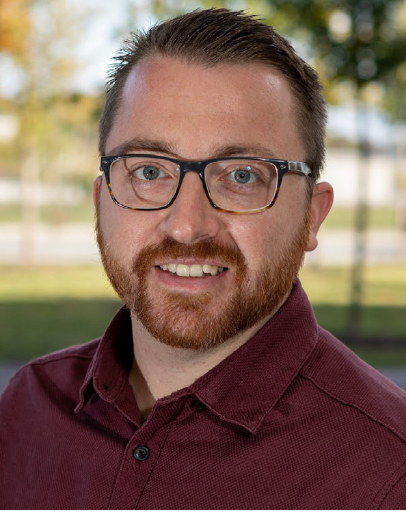}}]  {\textbf{JORGE QUEROL}} (S’13–M’18) was born in
Forcall, Castelló, Spain, in 1987. He
received the B.Sc. (+5) degree in
telecommunication engineering, the
M.Sc. degree in electronics engineering,
the M.Sc. degree in photonics, and the
Ph.D. degree (Cum Laude) in signal
processing and communications from the
Universitat Politècnica de Catalunya - BarcelonaTech (UPC),
Barcelona, Spain, in 2011, 2012, 2013 and 2018 respectively.
His research interests include Software Defined Radios (SDR),
real-time signal processing, satellite communications, 5G nonterrestrial
networks, satellite navigation, and remote sensing.
His Ph.D. thesis was devoted to the development of novel antijamming
and counter-interference systems for Global
Navigation Satellite Systems (GNSS), GNSS-Reflectometry,
and microwave radiometry. One of his outstanding
achievements was the development of a real-time standalone
pre-correlation mitigation system for GNSS, named FENIX, in
a customized Software Defined Radio (SDR) platform. FENIX
was patented, licensed and commercialized by MITIC
Solutions, a UPC spin-off company.
Since 2018, he is Research Scientist at the SIGCOM research
group of the Interdisciplinary Centre for Security, Reliability,
and Trust (SnT) of the University of Luxembourg,
Luxembourg. He is involved in several ESA and
Luxembourgish national research projects dealing with signal
processing and satellite communications.
He received the best academic record award of the year in
Electronics Engineering at UPC in 2012, the first prize of the
European Satellite Navigation Competition (ESNC) Barcelona
Challenge from the European GNSS Agency (GSA) in 2015,
the best innovative project of the Market Assessment Program
(MAP) of EADA business school in 2016, the award Isabel P.
Trabal from Fundació Caixa d’Enginyers for its quality
research during his Ph.D. in 2017, and the best Ph.D. thesis
award in remote sensing in Spain from the IEEE Geoscience
and Remote Sensing (GRSS) Spanish Chapter in 2019.
\end{IEEEbiography}

\begin{IEEEbiography}[{\includegraphics[width=1in,height=1.4in,clip,keepaspectratio]{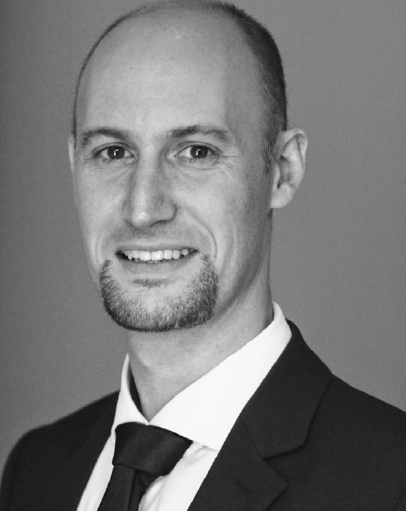}}]  {\textbf{SYMEON CHATZINOTAS}}  (Senior Member,
IEEE) is currently a Full Professor/Chief Scientist
I and the Head of the SIGCOM Research Group,
SnT, University of Luxembourg. He is coordinating
the research activities on communications
and networking, acting as a PI for more than
20 projects and main representative for 3GPP,
ETSI, and DVB. In the past, he has been a Visiting
Professor with the University of Parma, Italy, lecturing
on 5G wireless networks. He was involved
in numerous research and development projects for NCSR Demokritos,
CERTH Hellas and CCSR, and the University of Surrey. He has coauthored
more than 450 technical papers in refereed international journals,
conferences, and scientfic books. He was a co-recipient of the 2014 IEEE
Distinguished Contributions to Satellite Communications Award and the
Best Paper Awards at EURASIP JWCN, CROWNCOM, and ICSSC. He is
on the Editorial Board of the IEEE TRANSACTIONS ON COMMUNICATIONS, IEEE
OPEN JOURNAL OF VEHICULAR TECHNOLOGY, and the International Journal of
Satellite Communications and Networking.
\end{IEEEbiography}

\end{document}